\documentclass[a4paper,10pt]{article}
\usepackage{graphicx}
\usepackage{t1enc}

\begin{document}

\title{Wide-field stellar photometry in Piwnice Observatory}%
\author{Gracjan Maciejewski\\Centrum Astronomii UMK, ul. Gagarina 11, 87-100 Toru\'n\\gm@astri.uni.torun.pl}%
\maketitle

\textit{Abstract}: In this paper research projects based on the wide-field CCD photometry performed in Piwnice Observatory are discussed. The used telescopes, as well as dedicated software pipeline for data reduction are presented. The prospects for collaboration between Polish and Bulgarian institutes in the field of wide-field photometry are also discussed.
 

\section{Introduction}

Thanks to availability of large CCD detectors at a reasonable price and rapid decrease of the cost of the computing power needed for data reduction many massive survey projects have been initiated for the last decade. The way projects are performed depends significantly on their scientific goals. The main topics of these studies are: gravitational microlensing events, variable stars, optical
counterparts of gamma-ray bursts, planetary transits, supernovae, near-Earth objects, asteroids and comets. Nowadays even a very simple telescope equipped with a low-cost commercial CCD detector and a telephoto lens can perform many interesting scientific programs (Paczy\'nski 1997). The All Sky Automated Survey (Pojma\'nski 1997) is an excellent example of such a low-end project discovering thousands of new
variable stars.

The availability of large CCD detectors allowed to reactivate Schmidt cameras and to use their capability of deep imaging wide and deep fields. These instruments are for example excellent tools for open cluster surveys (e.g. Sharma et al. 2006, Maciejewski \& Niedzielski 2007). For many of them (half of open clusters  recorded in catalogues (Dias et al. 2002)) basic parameters, such as age, reddening, and distance, still remain unknown. The effects of an evaporation of low-mass members and a mass segregation, predicted by the dynamics of clusters, are still unstudied for the majority of clusters. Thanks to Schmidt cameras' capabilities, it is possible to obtain a complete picture of a cluster, not only its most dense center but also the expanded and sparse coronal region.

In this review two instruments -- the Small Camera and the Schmidt-Cassegrain Telescope, located in the Astronomical Observatory of the Nicolaus Copernicus University in Piwnice near Torun, Poland -- are presented. Their capabilities for wide-field stellar photometry are discussed in Sec.~\ref{sekcja2} and \ref{sekcja3}. In Sec. \ref{sekcja4} the universal software pipeline for the automatic data reduction is presented.

\section{Wide-field photometry with a very small telescope}\label{sekcja2}

An observational project called Semi-Automatic Variability Search\footnote{http://www.astri.uni.torun.pl/\~{}gm/SAVS} (SAVS, Niedzielski et al. 2003, Maciejewski \& Niedzielski 2005) has been operated in the Piwnice Observatory since 2002. It uses the existing infrastructure of the observatory and very simple, low-end, dedicated hardware. It is composed of CCD camera, simple optics, camera mount and a control computer. The main goal of the survey is to discover new bright variable stars in the northern hemisphere. Despite of numerous surveys performed last decade, thousands of bright ($V<14$ mag) variable stars still remain undiscovered.

Both SBIG ST-7XE and ST-8XE commercial CCD cameras are used as detectors. The optical system is constituted by an achromatic telephoto Telezenitar-M 135/2.8. Such a configuration, depending on the CCD camera in use, results in a field of view of 3 $\times$ 2 or 6 $\times$ 4 degree with a scale of 13.8 arcsec/pixel. The cameras are equipped with filter wheels containing the Johnson-Cousins UBVRI filter set, closely approximating the standard system. The observations are mainly collected in V band. With the 240 s exposures stars brighter than $V=14$ mag can be monitored. 

Meade LX200 telescope is used as a mount for positioning the camera at the required coordinates and tracking the sky rotation. The CCD camera with the lens is attached at the top of the telescope tube. The instrument, as well as the dome are controlled by a dedicated computer. The custom-made software runs in an automatic mode performing observations of selected fields. During daylight the telescope is set to a standby mode. After twilight the tracking is switched on and regular observations start in remote operation mode. The instrument is positioned at required coordinates of the first program field and then exposure starts. While an acquired CCD image is being downloaded, compressed on-the-fly and saved on a local hard disk, the telescope is already slewing towards following target. 
The list of targets loops during a night. Every field is observed with a priority and at a frequency set by an observer. Observations usually last till dawn. The role of observer was reduced to minimum. The human interaction is required for opening and closing dome's slit, inspecting acquired images and rejecting those of bad quality, archiving collected data and finally running a software pipeline for data reduction.

\begin{figure*}
\begin{center}
\includegraphics[width=0.5\textwidth]{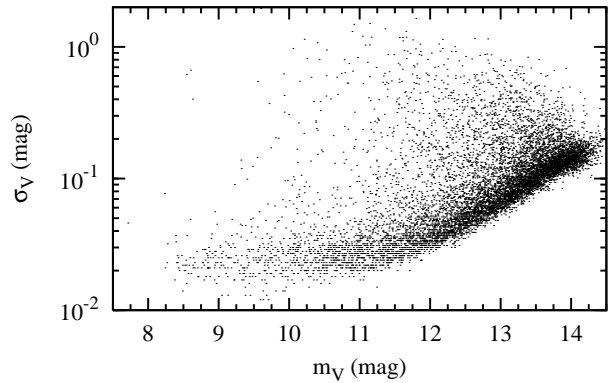}
\end{center}
\caption{Magnitude error vs. magnitude diagram for 10000 stars monitored by the SAVS project.}
\label{figure2}
\end{figure*}

Between 2002-2006 more than 200 variables of various types were discovered. For more than 120 of them the results have been published in numerous papers (e.g. Maciejewski et al. 2005). The efficiency of the survey is 1 new variable for about 500 monitored stars. The accuracy of collected photometry in V band is displayed in Fig.~\ref{figure2}. For stars blighter than $V=11$ mag, the typical error is lower than 0.03 mag. For fainter stars it increases as expected, reaching about 0.15 mag for stars of $V=14$ mag.

The SAVS project is currently used by students for educational purposes, as well as for searching new variables in selected fields and monitoring some interesting bright variables.

\section{Deep-field photometry with the 90-cm Schmidt camera}\label{sekcja3}

The Schmidt-Cassegrain Telescope, located in our observatory, was upgraded in 2005 what allowed us to use it in the Schmidt camera mode for wide and deep field imaging. Its main mirror diameter is 90 cm and focal length 180 cm what makes it very similar to the Schmidt telescope located at the National Astronomical Observatory (NAO) at Rozhen. The telescope is equipped with a correction plate with a 60 cm diameter and a field-flattening lens mounted near the focal plane. SBIG STL-11000 CCD camera with a KAI-11000M CCD detector (4008 $\times$ 2672 pixels $\times$ 9 $\mu$m) is used as a detector. The camera is equipped with a filter wheel containing standard UBVR Johnson-Cousins filters. The field of view of the instrument is 72 arcmin in declination and 48 arcmin in right ascension with the scale of 1.08 arcsec per pixel. The instrument is capable of reaching stars as faint as 19.5 mag in $V$ with 10 minute exposures. Due to considerable seeing (FWHM 5--6 arcsec), typical for our observing location, the $2 \times 2$  hardware binning is used to increase the signal-to-noise ratio.

Our Schmidt camera was found to be an excellent tool for complex studies of open clusters of diameter between 5 and 20 arcmin. The recently performed Open Cluster Survey\footnote{http://www.astri.uni.torun.pl/\~{}gm/OCS} (OCS, Maciejewski \& Niedzielski 2007) is a project which aim is to determine basic astrophysical parameters describing structure, dynamics and evolution state of selected open clusters. The statistical relations between some cluster parameter were investigated and compared with theoretical predictions. 

\begin{figure*}
\begin{center}
\includegraphics[width=0.5\textwidth]{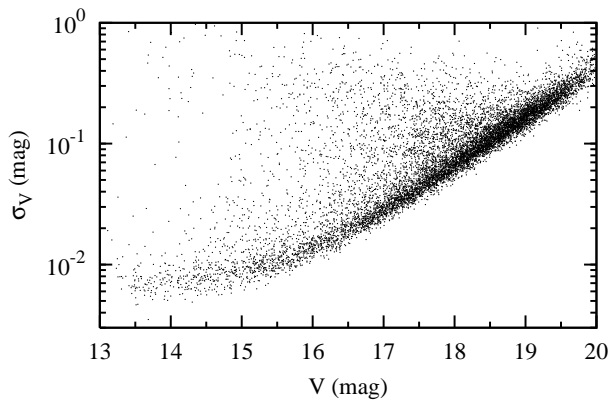}
\end{center}
\caption{Magnitude error vs. magnitude diagram obtained for stars monitored with the Schmidt-Cassegrain Telescope.}
\label{figure3}
\end{figure*}

The instrument is also used for searching variable stars in fields of selected open clusters. This program is performed in the collaboration with our Bulgarian colleagues who use 70/50-cm Schmidt telescope at NAO at Rozhen. In every monitored field tens of variables were discovered. Some of them belonging to open clusters with the highest probability, are especially valuable because their light curve analysis provides additional information on astrophysical parameters of star cluster. The accuracy of the photometry obtained in the survey is below 0.01 mag for stars brighter than $V=15$ mag with 10 minute exposures. As shown in Fig.~\ref{figure3}, we achieved errors below 0.1 mag for stars brighter than about $V=18$.

\section{Data reduction pipeline}\label{sekcja4}

A dedicated software pipeline was developed for the SAVS survey purposes. It is an easy-to-use and intuitive software which was created for semi-automatic reduction, analysis of a large amount of CCD images, and detection of new variable stars. The package performs standard reduction steps on raw CCD frames by removing instrumental effects such as dark current and flat-field. In an automatic way it detects all stars in every frame and conducts precise astrometry for them by transforming instrumental coordinates into equatorial. Differential aperture photometry of all objects detected in a CCD frame is performed against selected comparisons. The code also archives, manages and analyzes collected data. In a fully automatic way it joins data points from single observations into extended databases containing all reduced observations. 

The candidates for new variable stars are selected using the analysis of variance (ANOVA) method (Schwarzenberg-Czerny 1996). The search for candidates is performed on all detected stars what allows us to find variables with very small amplitudes, comparable to our photometric precision. Moreover, this method reduces the number of false candidates which variability is actually generated for example by hot pixels or cosmic-ray events.

The software is under permanent development. It is used for complete reduction of images obtained with our Schmidt-Cassegrain Telescope, as well as exposures acquired from 70/50-cm Schmidt telescope at NAO at Rozhen. The pipeline is available at the SAVS sky survey web site. 
 
\section{Conclusion}

In this review photometric surveys performed in Piwnice Observatory were summarized. The wide-field photometry with small telescope such as the Small Camera used for SAVS survey, is a powerful tool for educational purposes and can still bring some valuable scientific results. The advantages of Schmidt camera can be exploited in deep and wide field investigations of open clusters. Due to congenial scientific programs and similarity of instruments located at Piwnice Observatory and at NAO at Rozhen -- the 90/60 cm Schmidt-Cassegrain Telescope and 70/50 Schmidt Telescope, respectively, fruitful cooperation between Polish and Bulgarian scientists is foreseen. 

\textit{Acknowledgements.} This paper is a result of PAS/BAS exchange and joint research project \textit{Spectral and photometric studies of variable stars}. This research is also supported by UMK grant 369-A and Polish Ministry of Science and Higher Education grant 1P02D00730.

\end{document}